\documentstyle[12pt,a4,axodraw,epsfig]{article}
\newcommand\slv{v\kern-5pt\raise1pt\hbox{$\scriptstyle/$}\kern1pt}

\newcommand\be{\begin{equation}}
\newcommand\ee{\end{equation}}
\newcommand{\eps}{\varepsilon}
\def\bq{\begin{eqnarray}}
\def\eq{\end{eqnarray}}

\newlength{\dinwidth}                                                           
\newlength{\dinmargin}                                                          
\begin{document}
\thispagestyle{empty}
\begin{flushright}
NIKHEF-98-033\\
\end{flushright}
\vspace{0.5cm}
\begin{center}
{\Large \bf Reduction of multi-leg loop integrals}\\[.3cm]
\vspace{1.7cm}
$\mbox{{\sc \bf S.~Weinzierl}}^\ast$ \\[1cm]
\begin{center} \em 
NIKHEF, P.O. Box 41882, NL - 1009 DB Amsterdam, The Netherlands
\end{center}\end{center}
\vspace{2cm}

\begin{abstract}\noindent
{I give an efficient algorithm for the reduction of multi-leg one-loop
integrals of rank one. The method combines the basic ideas of the spinor algebra 
approach with the dual vector
approach and is applicable to box integrals or higher point integrals, if at least one external leg is massless.
This method does not introduce Gram determinants
in the denominator. It completes an algorithm recently given by R. Pittau.}    
\end{abstract}
\vspace{3cm}

\vspace*{\fill}

\noindent 
$^\ast${\small email address : stefanw@nikhef.nl}

\newpage

\section{Introduction}

Next-to-leading order calculations have become nowadays the standard
for perturbative calculations in high energy physics.
One of the major obstacles is the tedious reduction of tensor
loop integrals (e.g. integrals where the loop momentum appears in the
numerator) to standard scalar integrals. General algorithm have been
known for a long time (like the Passarino-Veltman algorithm \cite{ref1}, \cite{ref1a} or
the Feynman parameter space technique \cite{ref2}). Although they solve the
problem in principle, the resulting expressions are usually quite long
and involve the appearence of unphysical singularities in the form of
Gram determinants in the denominator. These singularities usually drop
out only after combination of many different structured terms.\\
A variety of alternative approaches exists:
J.M. Campbell, E.W.N. Glover and D.J. Miller \cite{ref4a} introduced
scalar integral functions in higher dimensions and absorbed the Gram
determinants in a linear combination of these.
D.B. Melrose \cite{ref2a} and independently W.L. van Neerven and J.A.M. Vermaseren \cite{ref3}
have used the fact, that in four dimensions there are maximally four
linearly independent vectors, to derive a relation between a scalar
pentagon integral and scalar box integrals. 
The derivation of van Neerven and Vermaseren was based
on dual vectors. The dual vector approach was further used by
G.J. van Oldenborgh and J.A.M. Vermaseren \cite{ref3a} and A. Signer \cite{ref5a}.
Recently R. Pittau \cite{ref4} has given
an algorithm, which makes use of the spinor algebra, to reduce tensor
integrals to scalar and rank one integrals.
In this paper I give a formula which allows the reduction
of the remaining rank one integrals to scalar integrals. It combines the 
basic ideas 
of the spinor
algebra approach and the dual vector approach. This formula completes
the algorithm given by R. Pittau.
\\
This paper is organized as follows: The following section introduces the
necesarry notation. In section 3 I derive the basic
formula. The complete algorithm is given in section 4. An example is worked
out in section 5. Section 6 contains the conclusions.

\section{Notation}

For an introduction to spinor products I refer to \cite{lecture}.
A typical one-loop integral has the form
\bq
I & = & \int \frac{d^D k}{(2 \pi)^D} \frac{\langle p_1 \pm| ... k\!\!\!/ ... |q_1 \pm \rangle
... \langle p_m \pm| ... k\!\!\!/ ... |q_m \pm \rangle}
{N_1 N_2 ... N_n}
\eq
where the $p_i$ and the $q_i$ are external massless momenta and the $N_i$ denote the internal
propagators. The loop momentum $k$ is allowed to appear more than once in each sandwich
$\langle p_i \pm| ... k\!\!\!/ ... |q_i \pm \rangle$.
This form is general:
After the decay of (possible) intermediate heavy particles all external particles can be taken
to be
gluons, photons or light fermions. The polarization vectors of gluons and photons
are expressed in the above form using the spinor helicity method \cite{ref5}:
\bq
\eps_\mu^+(p,q) = \frac{\langle q- | \gamma_\mu | p- \rangle}{\sqrt{2} \langle q p \rangle}, &&
\eps_\mu^-(p,q) = \frac{\langle q+ | \gamma_\mu | p+ \rangle}{\sqrt{2} \left[ p  q \right]} ,
\eq
where $q$ is an arbitrary null reference momentum.
It will be usefull to introduce a slight generalization of spinor products:
Consider first elements of the form $a\!\!\!/ = a^\mu \gamma_\mu$. They form an algebra,
which I will denote by $A$.
It is convenient to think of the space of ket-spinors generated by the spinors
obtained from all external momenta as an left $A$-module $V$. The dual
space of bra-spinors is denoted by $V^\ast$ and is a right $A$-module.
The spaces $V$ and $V^\ast$ decompose into a direct sum of subspaces of
positive and negative helicity:
\bq
V = V_+ \oplus V_- && V^\ast = V^\ast_+ \oplus V^\ast_-
\eq
Non-vanishing spinor products are only obtained between elements of $V^\ast_+$ and $V_-$ or
between elements of $V^\ast_-$ and $V_+$. They are denoted by
\bq
\left[ a b \right] = \langle a + | b - \rangle & & \langle a b \rangle = \langle a - | b + \rangle
\eq
where $\langle a \pm | \in V^\ast_{\pm}$ and $| b \pm \rangle \in V_\pm$.
Multiplication of a spinor with an element of $A$ changes the helicity, e.g.
if $|p+ \rangle \in V_+$ and $a\!\!\!/ \in A$ then $(a\!\!\!/ | p+ \rangle ) \in V_-$.
It is convenient to define a ``degree'' of these objects.
Assume that the spinor $| q \rangle $ corresponds to the external momentum $q$ and that
\bq
| p \rangle & = & a\!\!\!/_n ... a\!\!\!/_1 | q \rangle .
\eq
Then we define $\mbox{deg}( | q \rangle ) = 0$ and
\bq
\mbox{deg} \left( |p \rangle \right) & = & n.
\eq
A similar definition applies to bra-spinors. Changing the helicity or flipping a spinor from
bra to ket (or vice versa) does not change the degree. It follows that for arbitrary spinors
$| p \rangle $, $| q \rangle $, $| k \rangle $ and $| j \rangle $
\begin{eqnarray}
 \langle q p \rangle & = & - (-1)^{deg \; p} (-1)^{deg \; q} \langle p q \rangle, \nonumber \\
 \left[  q p  \right] & = & - (-1)^{deg \; p} (-1)^{deg \; q} [  p q  ], \nonumber \\
 \langle p q \rangle \langle k j \rangle & = & \langle p j \rangle \langle k q \rangle 
+ (-1)^{deg \; q} (-1)^{deg \;k} \langle p k \rangle \langle q j \rangle , \nonumber \\
 \left[ p q \right] [ k j ] & = & [ p j ] [ k q ] + (-1)^{deg \; q} (-1)^{deg \;k} [ p k ] [ q j ]. 
\end{eqnarray}
For $\mbox{deg} \; p = \mbox{deg} \; q = \mbox{deg} \; k = 0$ these formulae coincide with the usual ones for spinor products.\\
\\
I take the sign of the antisymmetric tensor as
$
\eps_{0123} = + 1
$.
It is useful to introduce the following short-hand notation:
$
\eps(q_a,q_b,q_c,q_d) = 4 i \eps_{\mu\nu\rho\sigma} q_a^\mu q_b^\nu q_c^\rho q_d^\sigma
$

\section{Derivation of the basic formula}

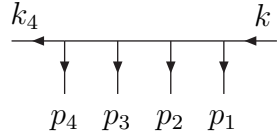
\begin{figure}
\begin{center}
\begin{picture}(100,100)(0,0)
\ArrowLine(100,70)(80,70)
\Line(80,70)(20,70)
\ArrowLine(20,70)(0,70)
\ArrowLine(20,70)(20,50)
\ArrowLine(40,70)(40,50)
\ArrowLine(60,70)(60,50)
\ArrowLine(80,70)(80,50)
\Text(20,40)[c]{$p_4$}
\Text(40,40)[c]{$p_3$}
\Text(60,40)[c]{$p_2$}
\Text(80,40)[c]{$p_1$}
\Text(95,80)[c]{$k$}
\Text(5,80)[c]{$k_4$}
\end{picture}
\caption{\label{figure1}
A subgraph of a one-loop diagram. The loop momentum is flowing through $k$ and $k_4$, the $p_i$'s 
are assumed to be the external momenta.}
\end{center}
\end{figure}

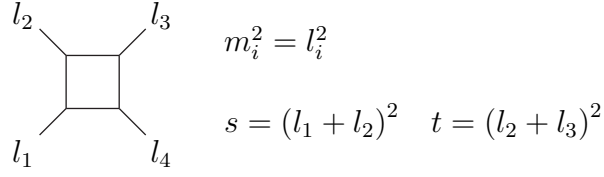
\begin{figure}
\begin{center}
\begin{picture}(100,100)(60,0)
\Line(40,40)(60,40)
\Line(40,40)(40,60)
\Line(60,60)(40,60)
\Line(60,60)(60,40)
\Line(40,40)(30,30)
\Line(40,60)(30,70)
\Line(60,40)(70,30)
\Line(60,60)(70,70)
\Text(28,28)[tr]{$l_1$}
\Text(28,72)[br]{$l_2$}
\Text(72,72)[bl]{$l_3$}
\Text(72,28)[tl]{$l_4$}
\Text(100,65)[l]{$m_i^2 = l_i^2$}
\Text(100,35)[l]{$s = \left( l_1 + l_2 \right)^2 \;\;\; t = \left( l_2 + l_3 \right)^2 $}
\end{picture}
\caption{\label{figure2} The labelling of a box integral with external momenta $l_i$.}
\end{center}
\end{figure}

Combining the basic ideas of reduction methods based on dual vectors and on the spinor algebra, I derive
a formula, which allows an efficient reduction of higher point integrals.\\
The kinematics are shown in fig.\ref{figure1}.
The $p_i$ are supposed to be external momenta whereas $k$ is taken as a loop momentum. I assume that the $p_i$ are linearly independent. This means that the method
applies to pentagon or higher point integrals. A similar formula for box integrals
is given later on. Let us introduce the notation
\begin{eqnarray}
q_i & = & \sum\limits_{j=1}^i p_i, \nonumber \\
k_i & = & k-q_i.
\end{eqnarray}
It is convenient to set $k_0=k$ and $p_5=-q_4$, so that one basically always deals with
five external momenta $p_1,...,p_5$, even for hexagon or higher point integrals.
We have
\begin{eqnarray}
2 k \cdot q_i & = & k^2 - k_i^2 + q_i^2, \nonumber \\
2 k \cdot p_i & = & k_{i-1}^2 - k_i^2 + q_i^2 - q_{i-1}^2.
\end{eqnarray}
It is convenient to work in a scheme where the Dirac matrices are in four dimensions (like the four-dimensional
helicity scheme \cite{FDH}).
If the integral is finite we can take all vectors in four dimensions. The case of a divergent integral 
needs some special care:
I follow the approach by G. Mahlon \cite{eps1} and split the loop momentum into a four-dimensional part and a $(-2\varepsilon)$-dimensional part
\begin{eqnarray}
k^{(D)} & = & k^{(4)} + k^{(-2\varepsilon)}.
\end{eqnarray}
It is further assumed that a four-dimensional vector is orthogonal to a $(-2 \eps)$-dimensional one.
I start from the dual vectors written as
\begin{eqnarray}
v_\mu^a & = & \frac{1}{6} \eps^{abcd} 4 i \eps_{\mu\nu\rho\sigma} q_b^\nu q_c^\rho q_d^\sigma,
\end{eqnarray}
and the Schouten identity
\begin{eqnarray}
\label{schouten}
\eps(q_1,q_2,q_3,q_4) \left(k^{(4)} \right)^2 & = & \sum\limits_{a=1}^4 \left( k \cdot v^a \right) \left( k \cdot q_a \right).
\end{eqnarray}
This equation is related to four dimensions, therefore $k^{(4)}$ appears on the
left hand side. Since $q_a$ and $v_a$ are four-dimensional objects, the scalar
products of $k^{(-2\eps)}$ with $q_a$ or $v_a$ vanish and I dropped therefore the superscripts.
Now 
\begin{eqnarray}
k \cdot v^a & = & \pm \frac{1}{6} \eps^{abcd} \left( 2 \mbox{Tr}_{\pm} k\!\!\!/ q\!\!\!/_b q\!\!\!/_c q\!\!\!/_d - \mbox{Tr} k\!\!\!/ q\!\!\!/_b q\!\!\!/_c q\!\!\!/_d
\right) \nonumber \\
& = & \pm \frac{1}{6} \eps^{abcd} 2 \mbox{Tr}_{\pm} k\!\!\!/ q\!\!\!/_b q\!\!\!/_c q\!\!\!/_d
\end{eqnarray}
where $\mbox{Tr}_\pm$ denotes a trace of Dirac matrices with the insertion of a projection
operator $\frac{1}{2} \left( 1 \pm \gamma_5 \right)$. 
Since
\begin{eqnarray}
\mbox{Tr} (k\!\!\!/ q\!\!\!/_b q\!\!\!/_c q\!\!\!/_d) & = & (2 k \cdot q_b) (2 q_c \cdot q_d ) - (2 k \cdot q_c) (2 q_b \cdot q_d ) + (2 k \cdot q_d) (2 q_b \cdot q_c ) 
\end{eqnarray}
and each terms has one scalar product symmetric in two indices, these terms drop out when contracted with the antisymmetric tensor. Working out the r.h.s of eq. (\ref{schouten}) yields
\bq
\left( k \cdot v^a \right) \left( k \cdot q_a \right) & = & 
\pm \left( 2 k \cdot q_1 \mbox{Tr}_\pm k\!\!\!/ q\!\!\!/_2 q\!\!\!/_3 q\!\!\!/_4
          -2 k \cdot q_2 \mbox{Tr}_\pm k\!\!\!/ q\!\!\!/_1 q\!\!\!/_3 q\!\!\!/_4 \right. \nonumber \\
& & \left.          +2 k \cdot q_3 \mbox{Tr}_\pm k\!\!\!/ q\!\!\!/_1 q\!\!\!/_2 q\!\!\!/_4
          -2 k \cdot q_4 \mbox{Tr}_\pm k\!\!\!/ q\!\!\!/_1 q\!\!\!/_2 q\!\!\!/_3
    \right) 
\eq
Writing $2 k \cdot q_a = k^2 - k_a^2 + q_a^2$, one sees that the first two terms cancel propagators,
whereas the last term yields a contribution of
\begin{eqnarray}
\label{arranged}
\lefteqn{\pm \left(
q_1^2 \mbox{Tr}_\pm k\!\!\!/ q\!\!\!/_2 q\!\!\!/_3 q\!\!\!/_4
-q_2^2 \mbox{Tr}_\pm k\!\!\!/ q\!\!\!/_1 q\!\!\!/_3 q\!\!\!/_4
+q_3^2 \mbox{Tr}_\pm k\!\!\!/ q\!\!\!/_1 q\!\!\!/_2 q\!\!\!/_4
-q_4^2 \mbox{Tr}_\pm k\!\!\!/ q\!\!\!/_1 q\!\!\!/_2 q\!\!\!/_3
\right) =} & & \nonumber \\
& =& \pm \left( \mbox{Tr}_\pm k\!\!\!/ p\!\!\!/_1 p\!\!\!/_2 p\!\!\!/_3 p\!\!\!/_4 p\!\!\!/_5 + q_1^2 q_3^2 (k_0^2 - k_4^2) 
+ q_1^2 q_4^2 (k_3^2 - k_2^2) + q_2^2 q_4^2 (k_1^2 - k_0^2 ) \right)
\end{eqnarray}
The r.h.s. of eq. (\ref{arranged}) has been arranged such that no constant term survives.
One finally obtains
\begin{eqnarray}
\label{res_eq}
\lefteqn{
\mbox{Tr}_\pm k\!\!\!/_0 p\!\!\!/_1 p\!\!\!/_2 p\!\!\!/_3 p\!\!\!/_4 p\!\!\!/_5 =  
- k_0^2 \mbox{Tr}_\mp  p\!\!\!/_1 p\!\!\!/_2 p\!\!\!/_3 p\!\!\!/_4 
- k_0^2 \mbox{Tr}_\pm k\!\!\!/_1 p\!\!\!/_2 p\!\!\!/_3 p\!\!\!/_4 } & &\nonumber \\
& & + k_1^2 \mbox{Tr}_\pm k\!\!\!/_0 (p\!\!\!/_1+p\!\!\!/_2) p\!\!\!/_3 p\!\!\!/_4 
 - k_2^2 \mbox{Tr}_\pm k\!\!\!/_0 p\!\!\!/_1 (p\!\!\!/_2+ p\!\!\!/_3) p\!\!\!/_4 
 + k_3^2 \mbox{Tr}_\pm k\!\!\!/_0 p\!\!\!/_1 p\!\!\!/_2 (p\!\!\!/_3+ p\!\!\!/_4) \nonumber \\
& & - k_4^2 \mbox{Tr}_\pm k\!\!\!/_0 p\!\!\!/_1 p\!\!\!/_2 p\!\!\!/_3 
- (k^{(-2\eps)})^2  \left( \mbox{Tr}_\pm p\!\!\!/_1 p\!\!\!/_2 p\!\!\!/_3 p\!\!\!/_4 - \mbox{Tr}_\mp p\!\!\!/_1 p\!\!\!/_2 p\!\!\!/_3 p\!\!\!/_4 \right).
\end{eqnarray}
This equation holds if the Dirac matrices and the external momenta are in four dimensions. The loop momentum
$k$ may be in $D$ dimensions, allowing the application also to divergent integrals.
A careful inspection of the basic equation (\ref{schouten}) shows that the term $\eps(q_1,q_2,q_3,q_4) \left(k^{(4)}\right)^2$
on the left-hand side has to be in four dimensions, whereas the manipulations on the right-hand side involved only
scalarproducts $2 k \cdot q_a = k^2 -k_a^2 + q_a^2$ which may be continued into $D$ dimensions as long as the external
momenta $q_a$ stay in four dimensions.
The last term of eq. (\ref{res_eq}) is a correction term, which takes care of
divergent integrals.
If the integral is finite, $k$ may be taken to be four dimensional and the last term vanishes.
It will turn out that the correction term is irrelevant in the application of eq. (\ref{res_eq}) to pentagon
or higher point integrals.
The equation above may be rewritten as :
\begin{eqnarray}
\label{res1}
\mbox{Tr}_{\pm} \left( k\!\!\!/ p\!\!\!/_1 p\!\!\!/_2 p\!\!\!/_3 p\!\!\!/_4 p\!\!\!/_5 \right) & = & - \mbox{Tr}_{\mp}\left(p\!\!\!/_1 p\!\!\!/_2 p\!\!\!/_3 p\!\!\!/_4 \right) \left( N_0 + M_0^2 \right) \nonumber \\
& & - \frac{1}{2} \left( B_0 \pm \varepsilon(k-p_1,p_2,p_3,p_4) \right) \left( N_0 + M_0^2 \right) \nonumber \\
& & + \frac{1}{2} \left( B_1 \pm \varepsilon(k,p_1+p_2,p_3,p_4) \right) \left( N_1 + M_1^2 \right) \nonumber \\
& & - \frac{1}{2} \left( B_2 \pm \varepsilon(k,p_1,p_2+p_3,p_4) \right) \left( N_2 + M_2^2 \right) \nonumber \\
& & + \frac{1}{2} \left( B_3 \pm \varepsilon(k,p_1,p_2,p_3+p_4) \right) \left( N_3 + M_3^2 \right) \nonumber \\
& & - \frac{1}{2} \left( B_4 \pm \varepsilon(k,p_1,p_2,p_3) \right) \left( N_4 + M_4^2 \right) \nonumber \\
& & - (k^{(-2\eps)})^2  \left( \mbox{Tr}_\pm p\!\!\!/_1 p\!\!\!/_2 p\!\!\!/_3 p\!\!\!/_4 - \mbox{Tr}_\mp p\!\!\!/_1 p\!\!\!/_2 p\!\!\!/_3 p\!\!\!/_4 \right) \nonumber \\
\end{eqnarray}
with $N_i = k_i^2 - M_i^2$, where the internal propagators and masses are denoted by $N_i$ and $M_i$, respectively.
$B_i$ depends only on the pinched integral under consideration and is given by
\begin{eqnarray}
\label{box_const}
B_i & = & s t - m_1^2 m_3^2 - m_2^2 m_4^2.
\end{eqnarray}
Starting from the pentagon integral the internal propagator $i$ is first pinched.
This gives a box integral and the value of $B_i$ is calculated from that box integral
according to eq. (\ref{box_const}) and according to the kinematics shown in fig. \ref{figure2}.
Clearly the value of $B_i$ does not depend on the starting point for the labelling of the external
legs $l_j$ of the box integral.
Formula (\ref{box_const}) also applies 
to hexagon or higher point integrals, where the sum of the external
momenta from leg 5 to leg $n$ is identified with $p_5$.\\
For a tensor box-integral one can use
\begin{eqnarray}
\label{res2}
\mbox{Tr}_{\pm}\left(k\!\!\!/_0 p\!\!\!/_1 p\!\!\!/_2 p\!\!\!/_3 \right) & = & \frac{1}{2} \left( B \pm \varepsilon(k_0,p_1,p_2,p_3) \right) \nonumber \\
& & - \frac{1}{2} C_0 \left( N_0 + M_0^2 \right) 
 + \frac{1}{2} C_1 \left( N_1 + M_1^2 \right) \nonumber \\
& & - \frac{1}{2} C_2 \left( N_2 + M_2^2 \right) 
 + \frac{1}{2} C_3 \left( N_3 + M_3^2 \right) 
\end{eqnarray}
with
\begin{eqnarray}
C_i & = & m_1^2 + m_2^2 - m_3^2.
\end{eqnarray}
This equation is just obtained by expanding the trace. 
\begin{figure}
\begin{center}
\begin{picture}(100,100)(0,0)
\ArrowLine(60,40)(40,40)
\Line(40,40)(50,60)
\Line(60,40)(50,60)
\Line(40,40)(30,30)
\Line(60,40)(70,30)
\Line(50,60)(50,70)
\Text(28,28)[tr]{$m_1^2$}
\Text(54,70)[l]{$m_2^2$}
\Text(72,28)[tl]{$m_3^2$}
\Text(50,37)[t]{$j$}
\end{picture}
\caption{\label{figure3} The labelling of a triangle integral.}
\end{center}
\end{figure}
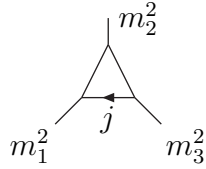
The orientation of the triangles depends on how the triangles are obtained from the box integral : If the loop
momentum $k_0$ is not pinched, $j=k_0$ and $m_1^2$ is the first external mass in the direction
of the flow of $j$. If $k_0$ is the pinched momentum, then $j=k_1$.
The kinematics are also shown in fig.\ref{figure3}.
Equation (\ref{res1}) and (\ref{res2}) are the basic equations for the reduction
of tensor loop integrals.
For box integrals the terms with the antisymmetric tensor in the numerator will vanish after integration.\\
\\
It remains to investigate under which condition the traces may be formed.
It turns out that it is always possible to form a trace $\mbox{Tr}_\pm k\!\!\!/ p\!\!\!/_1 p\!\!\!/_2 p\!\!\!/_3 p\!\!\!/_4 p\!\!\!/_5$ with
$p_5 = -p_1-p_2-p_3-p_4$, if at least one external leg is massless.
Terms like $ \langle q - | k\!\!\!/ | l - \rangle $, where k is the loop momentum, appear in 
the numerator of tensor integrals .
Using the Schouten identity twice, one can write:
\begin{eqnarray}
\label{trace}
\left[ i j \right] \langle i j \rangle \langle q - | k\!\!\!/ | l - \rangle & = & 
   - (-1)^{deg \; i} (-1)^{deg \; j} \langle q j \rangle [ j l ] \langle i - | k\!\!\!/ | i - \rangle \nonumber \\
                  &   & - (-1)^{deg \; i} (-1)^{deg \; j} \langle q i \rangle [ i l ] \langle j - | k\!\!\!/ | j - \rangle \nonumber \\
                  &   & + \langle q j \rangle [ i l ] \langle i - | k\!\!\!/ | j - \rangle \nonumber \\
                  &   & + (-1)^{deg \; i} (-1)^{deg \; j} \langle q i \rangle [ j l ] \langle i + | k\!\!\!/ | j + \rangle .
\end{eqnarray}
This will be the basic equation (together with suitable choices of $i$ and $j$) for forming the traces.
If a $n$-point integral ($n>4$) has at least one massless external leg, say $p_1$, one can choose
\begin{eqnarray}
| i- \rangle & = & | p_1 - \rangle , \nonumber \\
| j- \rangle & = & p\!\!\!/_5 p\!\!\!/_4 p\!\!\!/_3 p\!\!\!/_2 | p_1 - \rangle
\end{eqnarray}
where $p_5 = -p_4-p_3-p_2-p_1$. Of course, if there are more massless legs, simpler choices are possible. It is easily checked by induction that the terms $\langle i- | k\!\!\!/ | i- \rangle$ and
$\langle j- | k\!\!\!/ | j- \rangle$ always cancel propagators.

\section{The algorithm}

This gives us the following algorithm for the reduction of multi-leg loop integrals (e.g. at least
four external legs), which is applicable if at least one external leg is massless. Three-point and two-point
integrals are usually not so complicated and can be reduced using standard techniques.\\
\\
Step 1 consists in reducing tensor integrals
to rank one integrals and was given by R. Pittau \cite{ref4}: 
If there at least two massless legs, say
$p_1$ and $p_2$, one writes
\begin{eqnarray}
k\!\!\!/ & = & \frac{1}{2p_1 \cdot p_2} \left[ 
\left( 2 k \cdot p_2 \right) p\!\!\!/_1 + \left( 2 k \cdot p_1 \right) p\!\!\!/_2 - 
p\!\!\!/_1 k\!\!\!/ p\!\!\!/_2
-p\!\!\!/_2 k\!\!\!/ p\!\!\!/_1
\right].
\end{eqnarray}
The first two terms already cancel propagators, whereas a product of the last two terms can be rewritten as
\begin{eqnarray}
\langle 1- | k\!\!\!/ | 2- \rangle \langle 2- | k\!\!\!/ | 1- \rangle & = & \left( 2 k \cdot p_1 \right) \left( 2 k \cdot p_2 \right)
- \left( 2 p_1 \cdot p_2 \right) k^2, \nonumber \\
\langle 1- | k\!\!\!/ | 2- \rangle \langle 2+ | k\!\!\!/ | 1+ \rangle & = & \frac{1}{\langle 2- | 3 | 1- \rangle}
\left[ \left( 2 p_1 \cdot p_3 \right) \left( 2 p_2 \cdot k \right) \langle 1- | k\!\!\!/ | 2- \rangle \right. \nonumber \\
& & \left. -\left( 2 p_1 \cdot k \right) \langle 1- | p\!\!\!/_2 k\!\!\!/ p\!\!\!/_3 | 2- \rangle \right. \nonumber \\
& & \left. +\left( 2 p_3 \cdot k \right) \langle 1- | p\!\!\!/_2 k\!\!\!/ p\!\!\!/_1 | 2- \rangle \right. \nonumber \\
& & \left. + k^2  \langle 1- | p\!\!\!/_2 p\!\!\!/_3 p\!\!\!/_1 | 2- \rangle
\right],
\end{eqnarray} 
where $p_3$ is another (not necessarily massless) external leg. If there are no massless legs, or only one
massless leg, one can take linear combinations of the external momenta in order to form two massless
momenta \cite{ref4}. If all external momenta are massive this will re-introduce Gram determinants.\\
\\
In step 2 we reduce the remaining rank one integrals to scalar integrals by first forming the trace
with the help of equation (\ref{trace}). This can always be done if at least one
external leg is massless. Using equation (\ref{res1}) and equation (\ref{res2})
we then obtain scalar integrals, integrals with an antisymmetric tensor in the numerator and a correction term. For box integrals the terms with the antisymmetric
tensor in the numerator will vanish after integration. For higher point integrals one may rewrite
\bq
\eps(k,p_1,p_2,p_3) & = & \mbox{Tr}_+ \left( k\!\!\!/ p\!\!\!/_1 p\!\!\!/_2 p\!\!\!/_3 \right)
                         - \mbox{Tr}_- \left( k\!\!\!/ p\!\!\!/_1 p\!\!\!/_2 p\!\!\!/_3 \right)
\eq 
and iterate the procedure.\\
\\
Step 3 : The correction term is evaluated as follows 
\cite{eps2}:
\bq
\int \frac{d^{4-2\eps}k}{(2 \pi)^{4-2\eps}} \frac{\left( k^{(-2\eps)}\right)^2}{N_1 N_2 ... N_n}
& = &
\eps (4 \pi) \int \frac{d^{6-2\eps}k}{(2 \pi)^{6-2\eps}} \frac{1}{N_1 N_2 ... N_n},
\eq
The correction term is proportional to the scalar integral in $D = 6 - 2 \eps$ dimensions.
These integrals are finite in the limit $\eps \rightarrow 0$ for $n \ge 4$ \cite{ref2} and the correction term
does therefore not contribute to order $O(\eps^0)$.

\section{An example}

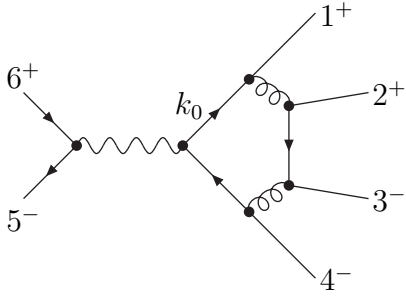
\begin{figure}
\begin{center}
\begin{picture}(100,140)(0,0)
\Vertex(20,50){2}
\ArrowLine(0,70)(20,50)
\ArrowLine(20,50)(0,30)
\Photon(20,50)(60,50){3}{4}
\Vertex(60,50){2}
\ArrowLine(60,50)(85,75)
\Line(85,75)(110,100)
\Line(110,0)(85,25)
\ArrowLine(85,25)(60,50)
\Vertex(85,75){2}
\Vertex(85,25){2}
\ArrowLine(100,65)(100,35)
\Vertex(100,65){2}
\Vertex(100,35){2}
\Gluon(85,75)(100,65){3}{2}
\Gluon(85,25)(100,35){3}{2}
\Line(100,65)(130,70)
\Line(100,35)(130,30)
\Text(0,72)[b]{$6^+$}
\Text(0,27)[t]{$5^-$}
\Text(112,100)[l]{$1^+$}
\Text(112,0)[l]{$4^-$}
\Text(132,70)[l]{$2^+$}
\Text(132,30)[l]{$3^-$}
\Text(68,62)[rb]{$k_0$}
\end{picture}
\caption{\label{figure4} The Feynman diagram with the helicity assignment. All external momenta are taken to be outgoing. }
\end{center}
\end{figure}

As an example I consider here a pentagon integral which was recently encountered in the 
one-loop calculation for $e^+ e^- \rightarrow q \bar{q} Q \bar{Q}$ \cite{amp1}. We want to evaluate the helicity diagram shown in fig. \ref{figure4}.
The diagram corresponds to the following integral:
\bq
I & = & - 8 \frac{\left[ 12 \right] \langle 34 \rangle}{s_{56}}
\int \frac{d^D k}{(2 \pi)^D} \frac{\langle 5- | k\!\!\!/_4 k\!\!\!/_2 k\!\!\!/_0 | 6- \rangle}{k_0^2 k_1^2 k_2^2 k_3^2 k_4^2}
\eq
The numerator is rewritten as
\bq
\label{num}
\langle 5- | k\!\!\!/_4 k\!\!\!/_2 k\!\!\!/_0 | 6- \rangle & = & 
- \langle 5- | \left( p\!\!\!/_3 + p\!\!\!/_4 \right) k\!\!\!/_2 \left( p\!\!\!/_1 + p\!\!\!/_2 \right) | 6 -\rangle \nonumber \\
& & + k_2^2 \langle 5- | k\!\!\!/_0 | 6- \rangle - k_2^2 \langle 5- | p\!\!\!/_3 + p\!\!\!/_4 | 6- \rangle
\eq
The first term on the r.h.s of eq. (\ref{num}) corresponds to a rank one pentagon integral, the
second term to a rank one box integral, whereas the last term already corresponds to a scalar box integral. The rank one pentagon integral is reduced as follows: Multiplying with 
$[23] \langle 23 \rangle$ and using eq. (\ref{trace}) we obtain:
\begin{eqnarray}
\lefteqn{[23] \langle23\rangle \langle5-|(p\!\!\!/_3+p\!\!\!/_4) k\!\!\!/_2 (p\!\!\!/_1+p\!\!\!/_2)|6-\rangle = } & & \nonumber \\
& = &  - \langle 6+ | 1+2 | 3+ \rangle \left[ 34 \right] \langle 45 \rangle \left( k_1^2 - k_2^2 \right) 
- \left[ 61 \right] \langle 12 \rangle \langle 2 + | 3+4 | 5+ \rangle \left( k_2^2 - k_3^2 \right) \nonumber \\
&   & + \frac{\langle 6+ | 1+2 | 3+ \rangle \langle 2 + | 3+4 | 5+ \rangle}{\langle 34 \rangle \langle 4+ | 5+6 | 1+ \rangle \left[ 12 \right]} \mbox{Tr}_+ \left( k\!\!\!/_2 p\!\!\!/_3 p\!\!\!/_4 \left(
p\!\!\!/_5 + p\!\!\!/_6 \right) p\!\!\!/_1 p\!\!\!/_2  \right) \nonumber \\
&   & + \frac{\left[ 61 \right] \langle 45 \rangle}{\langle 4- | 5+6 | 1- \rangle} \mbox{Tr}_- \left( k\!\!\!/_2 p\!\!\!/_3 p\!\!\!/_4 \left(
p\!\!\!/_5 + p\!\!\!/_6 \right) p\!\!\!/_1 p\!\!\!/_2 \right) .
\end{eqnarray}
According to eq. (\ref{res1}) the traces give after integration
\bq
\label{int1}
\lefteqn{
\int \frac{d^D k}{(2 \pi)^D} \frac{\mbox{Tr}_{\pm} \left( k\!\!\!/_2 p\!\!\!/_3 p\!\!\!/_4 \left(
p\!\!\!/_5 + p\!\!\!/_6 \right) p\!\!\!/_1 p\!\!\!/_2  \right)}{k_0^2 k_1^2 k_2^2 k_3^2 k_4^2} 
= \int \frac{d^D k}{(2 \pi)^D} \frac{1}{k_0^2 k_1^2 k_2^2 k_3^2 k_4^2} 
\left( - k_2^2 \mbox{Tr}_{\mp} \left( p\!\!\!/_3 p\!\!\!/_4 \left( p\!\!\!/_5 + p\!\!\!/_6 \right) p\!\!\!/_1\right) \right. 
} \nonumber \\
& & - \left. \frac{1}{2} \left( s_{123} s_{234} - s_{23} s_{56} \right) k_2^2 + \frac{1}{2} s_{12} s_{234} k_3^2 - \frac{1}{2} s_{12} s_{23} k_4^2 + \frac{1}{2} s_{23} s_{34} k_0^2 - \frac{1}{2} s_{123} s_{34} k_1^2 \right) \nonumber \\
& & + O(\eps)
\eq
The rank one box integral is evaluated as follow: After multiplication with
$\left[ 14 \right] \langle 14 \rangle$ we obtain
\bq
\lefteqn{\left[ 14 \right] \langle 14 \rangle
\langle 5- | k\!\!\!/_0 | 6- \rangle =
} \nonumber \\
& = & - \langle 54 \rangle \left[ 46 \right] \left( k_0^2 - k_1^2 \right)
    - \langle 51 \rangle \left[ 16 \right] \left( k_3^2 - k_4^2 + s_{56} - s_{123} \right) \nonumber \\
& & + \frac{\langle 54 \rangle \left[ 16 \right]}{\langle 4- | 2+3 | 1- \rangle } 
 \mbox{Tr}_+ \left( k\!\!\!/_0 p\!\!\!/_1 \left( p\!\!\!/_2 + p\!\!\!/_3 \right) p\!\!\!/_4 \right) \nonumber \\
& & + \frac{\langle 51 \rangle \left[ 46 \right]}{\langle 4+ | 2+3 | 1+ \rangle } 
 \mbox{Tr}_- \left( k\!\!\!/_0 p\!\!\!/_1 \left( p\!\!\!/_2 + p\!\!\!/_3 \right) p\!\!\!/_4 \right) 
\eq
The traces give after integration (using eq. (\ref{res2})):
\bq
\int \frac{d^D k}{(2 \pi)^D} \frac{\mbox{Tr}_{\pm} \left( k\!\!\!/_0 p\!\!\!/_1 \left( p\!\!\!/_2 + p\!\!\!/_3 \right) p\!\!\!/_4 \right)}{k_0^2 k_1^2 k_3^2 k_4^2}
& = & \frac{1}{2} \int \frac{d^D k}{(2 \pi)^D} \frac{1}{k_0^2 k_1^2 k_3^2 k_4^2} \left(
s_{123} s_{234} - s_{23} s_{56} 
\right. \nonumber \\
& & - \left( s_{23} - s_{234} \right) k_0^2 + \left( s_{56} - s_{123} \right) k_1^2 \nonumber \\
& & \left. - \left( s_{56} - s_{234} \right) k_3^2 + \left( s_{23} - s_{123} \right) k_4^2 \right)
\eq
This completes the reduction to scalar integrals.
In this example all propagators were massless. The extension to massive lines
is straightforward. If for example $k_2$ is replaced by a massive line, the algorithm yields
a combination of scalar boxes and a scalar pentagon integral.
The scalar pentagon integral can be expressed in terms of scalar boxes according to well-known
formulae \cite{ref2,ref3}.

\section{Conclusions}

In this paper, I have given a formula which allows the reduction of rank one pentagon or higher point 
integrals. If there is at least one massless external leg, this formula can always be applied.
This method does not introduce a Gram determinant in the denominator.
A corresponding formula for rank one box-integrals was also given. For rank one triangles or
bubbles conventional techniques are usually efficient enough.
The formulae in this paper complete an algorithm recently given by R. Pittau, and form together an
efficient algorithm for reducing tensor loop integrals.\\
\\
I would like to thank D.A. Kosower and J.A.M. Vermaseren for useful
discussions and for suggestions concerning the manuscript.


\begin{thebibliography}{99}

\bibitem{ref1} G.Passarino and M.Veltman, Nucl.Phys. B160, (1979), 151
\bibitem{ref1a} R.G. Stuart, Comp. Phys. Comm. 48, (1988), 367 \\
R.G. Stuart and A. G\'ongora-T., Comm. Phys. Comm. 56, (1990), 337
\bibitem{ref2} Z.Bern, L.Dixon and D.A.Kosower, Nucl.Phys. B412, (1994), 751
\bibitem{ref2a} D.B.Melrose, Nuovo Cimento 40A, (1965), 181
\bibitem{ref3} W.L.van Neerven and J.A.M.Vermaseren, Phys.Lett. 137B, (1984),241
\bibitem{ref3a} G.J. van Oldenborgh and J.A.M.Vermaseren, Z.Phys. C46, (1990), 425
\bibitem{ref4} R.Pittau, Comput.Phys.Commun. 104, (1997), 23 \\
               R.Pittau, Comput.Phys.Commun. 111, (1998), 48
\bibitem{ref4a} J.M.Campbell, E.W.N.Glover and D.J.Miller, Nucl. Phys. B498, (1997), 397
\bibitem{ref5a} A Signer, Ph.D. thesis, Diss. ETH Nr. 11143
\bibitem{ref5} P. De Causmaecker, R.Gastmans, W.Troost and T.T.Wu, Nucl.Phys. B206, (1982), 53 \\
               R.Kleiss and W.J.Stirling, Nucl.Phys. B262, (1985), 235 \\
               J.F.Gunion and Z.Kunzst, Phys.Lett. 161B, (1985), 333 \\
               Z.Xu, D.-H.Zhang and L.Chang, Nucl.Phys. B291, (1987), 392
\bibitem{lecture} M.Mangano and S.Parke, Phys.Rep. 200, (1991), 301 \\
                  L.Dixon, TASI 1995, hep-ph/9601359
\bibitem{FDH} Z.Bern and D.A.Kosower, Nucl.Phys. B379, (1992), 451
\bibitem{eps1} G.Mahlon, Phys.Rev. D49, (1994), 2197 \\
               G.Mahlon, Phys.Rev. D49, (1994), 4438
\bibitem{eps2} Z.Bern and A.G.Morgan, Nucl.Phys. B467, (1996), 479
\bibitem{amp1} Z.Bern, L.Dixon, D.A.Kosower and S.Weinzierl, Nucl.Phys. B489, (1997), 3
\end{thebibliography}
\end{document}